\documentclass[%
 reprint,
 amsmath,amssymb,
 aps,
 prd,
]{revtex4-2}

\usepackage{graphicx}
\usepackage{dcolumn}
\usepackage{bm}

\newcommand{\equref}[1]{Eq.~(\ref{#1})}
\newcommand{\figref}[1]{Fig.~\ref{#1}}
\newcommand{\tabref}[1]{Table ~\ref{#1}}
\newcommand{\secref}[1]{Sec.~\ref{#1}}

\newcommand{\vkep}{v_{\mathrm{kep}}}

\newcommand{\SMBH}{\mathrm{SMBH}}

\newcommand{\Msun}{\mathrm{M}_{\odot}}
\newcommand{\Edd}{\mathrm{Edd}}

\newcommand{\DBH}{\mathrm{DBH}}

\newcommand{\rHill}{r_{\mathrm{Hill}}}

\newcommand{\GW}{\mathrm{GW}}
\newcommand{\BH}{\mathrm{BH}}

\newcommand{\ini}{\mathrm{ini}}
\newcommand{\rmin}{\mathrm{in}}
\newcommand{\out}{\mathrm{out}}
\newcommand{\disk}{\mathrm{disk}}
\newcommand{\IMF}{\mathrm{IMF}}

\newcommand{\gas}{\mathrm{gas}}

\newcommand{\AGN}{\mathrm{AGN}}

\newcommand{\GDF}{\mathrm{GDF}}
\newcommand{\acc}{\mathrm{acc}}
\newcommand{\mig}{\mathrm{mig}}

\newcommand{\capt}{\mathrm{cap}}

\newcommand{\preexisit}{\mathrm{pre}}

\newcommand{\single}{\mathrm{single}}
\newcommand{\binary}{\mathrm{binary}}

\newcommand{\EMRI}{\mathrm{EMRI}}

\newcommand{\LISA}{\mathrm{LISA}}

\newcommand{\inc}{\rm{inc}}

\usepackage{xcolor,colortbl}

\begin{document}

\preprint{APS/123-QED}

\title{Recoil-regulated extreme mass-ratio inspirals in AGN disks}

\author{LingQin Xue}
\affiliation{Department of Physics, University of Florida, PO Box 118440, Gainesville, FL 32611-8440, USA}

\author{Zolt\'an Haiman}
\affiliation{Institute of Science and Technology Austria, Am Campus 1, Klosterneuburg 3400 Austria}
\affiliation{Department of Astronomy, Columbia University, MC 5246, 538 West 120th Street, New York, New York 10027, USA}
\affiliation{Department of Physics, Columbia University, MC 5255, 538 West 120th Street, New York, New York 10027, USA}

\author{Hiromichi Tagawa}
\affiliation{Shanghai Astronomical Observatory, Shanghai, 200030, People's Republic of China}

\author{Imre Bartos}
\email{imrebartos@ufl.edu}
\affiliation{Department of Physics, University of Florida, PO Box 118440, Gainesville, FL 32611-8440, USA}

\date{\today}

\begin{abstract}
Extreme mass-ratio inspirals (EMRIs) are among the primary targets of future space-based gravitational-wave observatories, such as LISA, TianQin, and Taiji. Active galactic nucleus (AGN) disks provide a gas-rich environment in which stellar-mass black holes can migrate toward central supermassive black holes and form EMRIs. Previous studies of this ``wet'' channel have largely neglected stellar interactions within the disk. Here we show that binary formation, hierarchical mergers, and recoil kicks fundamentally regulate wet EMRI formation in AGN disks. Using semi-analytical AGN disk models combined with Monte Carlo simulations across supermassive black hole masses of $10^5$--$10^7\,M_\odot$ and Eddington ratios of $10^{-3}$--1, we find that recoil kicks from mergers and  binary--single interactions repeatedly lift stellar-mass black holes out of the disk plane, temporarily interrupting migration and strongly suppressing EMRI formation in much of parameter space. Detectable EMRIs are therefore preferentially produced in young AGNs, typically within $\sim 10$--20\,Myr of disk formation, and often involve merger-grown secondary black holes. We predict LISA detection rates of $\sim 1$--30\,yr$^{-1}$, with the observable population dominated by low-mass AGNs and sensitive to the poorly constrained demographics of faint active nuclei. Our results identify stellar interactions as a key ingredient in the evolution of compact objects in AGN disks and show that future EMRI observations can probe both AGN disk physics and the low-mass AGN population.
\end{abstract}


\maketitle

\section{Introduction} \label{sec:intro}

The forthcoming generation of space-based gravitational-wave (GW) observatories, including LISA \cite{colpi2024lisa,baker2019laser}, TianQin \cite{luo2016tianqin,mei2021tianqin}, and Taiji \cite{Hu:2017mde,Ruan2020Taiji}, will open a new observational window in the mHz band, inaccessible to ground-based detectors. This low-frequency regime is expected to host diverse GW sources, including massive black hole binaries, Galactic binaries, and extreme mass-ratio inspirals (EMRIs) \cite{amaro2007intermediate,Berry:2019wgg,drasco2006gravitational}.
An EMRI consists of a supermassive black hole (SMBH) of $10^5$-$10^7\,\Msun$ and a compact secondary object, typically a stellar-mass black hole (sBH) or neutron star, with mass $\mathcal{O}(1$-$100)\,\rm{\Msun}$. As the secondary inspirals through GW emission, it spends months to years in the detector band, generating long-lived, information-rich signals \cite{barack2004lisa}. These encode the relativistic dynamics of the system, providing tests of general relativity in the strong-field regime \cite{barack2007using,tahura2024vacuum,cardenas2024testing} and probes of astrophysical and cosmological environments \cite{barausse2014can,barausse2015environmental,gair2017prospects,laghi2021gravitational,li2022probing,duque2024extreme,derdzinski2021evolution,derdzinski2019probing,derdzinski2023situ}.

Two primary formation channels are typically considered for EMRIs. In stellar-dynamical environments such as nuclear star clusters, EMRIs can form through mechanisms including loss-cone scattering of compact objects \cite{lightman1977distribution,amaro2018relativistic}, tidal separation of stellar binaries \cite{hills1988hyper,miller2005binary}, and tidal stripping of giant stars \cite{chen2011tidal,bode2014production,naoz2022combined,mazzolari2022extreme}. These processes collectively define the \emph{dry} EMRI channel. In contrast, gas-rich environments, such as active galactic nucleus (AGN) disks, enable an alternative \emph{wet} EMRI pathway, in which compact objects embedded in the disk undergo orbital migration driven by gravitational torques from the surrounding gas \cite{baruteau2010binaries,li2021orbital,li2022long,rowan2023black}. Over time, these torques can transport compact objects inward toward the central SMBH, facilitating EMRI formation in the presence of gas.

The two channels are expected to yield distinct orbital and spin properties. Gas-assisted migration in AGN disks typically produces systems with low eccentricities, small inclinations, and characteristic secondary-spin alignments \cite{cui2025secondary,sun2025probing,duque2025constraining,li2025extreme}, whereas stellar-dynamical captures generally result in moderately eccentric systems and randomly oriented spins \cite{levin2007starbursts,pan2021formation,derdzinski2023situ}. Distinguishing wet and dry EMRIs in future GW observations will therefore provide valuable insights into both AGN disk physics and stellar dynamics in galactic nuclei.

Several studies have explored the wet EMRI channel. \cite{pan2021wet,pan2021formation} employed a Fokker-Planck framework to show that wet EMRIs may constitute a substantial, potentially dominant, fraction of all EMRIs detectable by space-based GW observatories. \cite{pan2022mass} demonstrated that this channel efficiently produces EMRIs in the low-mass regime, enhancing their detectability for upcoming missions. \cite{derdzinski2023situ} estimated a wet EMRI formation rate of $\Gamma_{\rm EMRI} \sim 0$-$10^{-4}\,\mathrm{yr^{-1}}$ per AGN from stars, assuming the presence of migration traps. More recently, \cite{sun2025probing} compared the mass and orbital parameter distributions of wet and dry EMRIs, identifying potentially observable distinctions between the two formation pathways. However, previous studies have largely neglected the role of hierarchical mergers or stellar interactions among sBHs within AGN disks, despite growing evidence that AGN disks may efficiently produce repeated mergers and dynamically assembled compact-object binaries \citep{2019PhRvL.123r1101Y,2020ApJ...901L..34Y,2021MNRAS.507.3362T,2022Natur.603..237S}.

In this work, we investigate how stellar interactions and hierarchical mergers among sBHs shape the population of wet EMRIs. We employ a semi-analytical framework built on a standard AGN disk model, coupled with a minimal prescription for migration. By incorporating global distributions of SMBH masses and accretion rates (Eddington ratios), we estimate the detection rate and mass distribution of wet EMRIs for future space-based GW detectors. Furthermore, we explore the resulting secondary-mass distribution of detectable EMRIs, providing a first step toward linking AGN disk physics to the observational landscape expected from LISA, TianQin, and Taiji.

The paper is organized as follows. Section~\ref{sec:method} describes the semi-analytical framework. Section~\ref{sec:result} presents the simulation results and discussion. We summarize our conclusions in Section~\ref{sec:conclusion}.

\section{Method}
\label{sec:method}

We model the formation of wet EMRIs in AGNs hosting an accretion disks. The system consists of three basic ingredients: a central SMBH, a gaseous accretion disk, and a population of sBHs initially distributed around the SMBH. The sBH population includes both initially single objects and a fraction of pre-existing binaries. As the sBHs interact with the AGN disk, their inclinations can be damped, their orbits can
migrate radially, and they can grow through gas accretion or repeated mergers. Some of these objects eventually reach the LISA band and are identified as wet EMRIs.

Our calculation builds on the semi-analytic framework of \cite{xue2025determines}. The workflow proceeds in five steps. First, we specify the initial SMBH--sBH configuration and the radial range of the simulation. Second, we compute the AGN disk structure and the physical rates governing sBH evolution. Third, we evolve the statistical background distributions of disk-embedded single and binary sBHs, $\Sigma(r,M)$, throughout the AGN lifetime. Fourth, we stochastically generate individual sBHs using a Monte Carlo approach and follow their dynamical evolution in the AGN environment, yielding populations of wet EMRIs and binary-sBH mergers. Finally, we incorporate AGN demography into a cosmological formation-rate calculation.

Several aspects of the original framework are modified for the present wet-EMRI calculation, as described in the following subsections. In \secref{sec:ini}, we specify the initial sBH demography and simulation range, which are kept fixed throughout all simulations. In \secref{sec:AGN}, we describe the adopted self-gravitating AGN disk model of \cite{sirko2003spectral}, which maintains a constant accretion rate in the $\alpha$-disk and naturally connects the disk structure to the
observed Eddington-ratio distribution of AGNs. In \secref{sec:process}, we summarize the physical mechanisms governing sBH evolution in the disk. A major update in this part is the inclination-damping prescription, for which we use recent
hydrodynamical results \cite{whitehead2025hydrodynamic,rowan2025hydrodynamic}.
In \secref{sec:back}, we extend the statistical background-function method by including conversion between single and binary sBH populations, using a simplified but computationally efficient prescription motivated by the conclusions of \cite{xue2025determines}. In \secref{sec:individual}, we implement the updated background treatment in Monte Carlo simulations of individual sBH evolution. In
\secref{sec:SNR}, we assess the LISA detectability of EMRIs from AGNs hosting different SMBH and sBH masses by estimating their maximum observable redshift. Finally, in \secref{sec:cos}, we combine the AGN mass and Eddington-ratio distributions with the formation-rate calculation presented in \secref{sec:formation}.

\begin{table*}
\caption{\label{tab:para}%
The BH and stellar distribution parameters that remain fixed throughout this paper. All the parameters listed here are the fiducial parameters and equations from \cite{xue2025determines}.}

\begin{ruledtabular}
\begin{tabular}{lll}
\textrm{parameter symbol}&
\textrm{description}&
\textrm{value}\\
\colrule
$\beta_{\IMF}$ & initial mass function of sBH &2.35 \\
$M_{\BH,\min}$ & minimum mass of pre-existing sBH & $5\Msun$ \\
$M_{\BH,\max}$ & maximum mass of pre-existing sBH & $15\Msun$\\
$\gamma_{\rho}$ & index of radial distribution function of pre-existing sBH & $0.0$\\
$N_{\BH,\ini}$ & total number of pre-existing sBH & $M_{\SMBH}/(20\Msun)$\\
$r_{\BH,\out}$ & maximum radius of pre-existing sBH & $GM_{\SMBH}/\sigma^2$ \\
$\sigma$ & velocity dispersion & $265\;\mathrm{km/s}\,(M_{\SMBH}/10^8\;M_{\odot})^{1/4}$ \\
$\beta_{v}$ & velocity dispersion parameter of pre-existing sBH & 0.2\\
$f_{\preexisit}$ & fraction of pre-existing sBH binaries&  0.15\\
$R_{\max}$ & maximum separation of pre-existing sBH binaries &  $10^5\rm{R_{\odot}}$\\
$\Gamma_{\Edd}$ & maximum accretion rate with respect to Eddington accretion rate & 1\\
$\eta_{c}$ & radiation efficiency of sBH accretion & 0.1\\
$\epsilon$ & radiation efficiency of SMBH accretion & 0.1\\
$\alpha$ & viscosity parameter & 0.01\\
$r_{\rmin}$ & minimum radius of simulation & $r_{\LISA}$ corresponds to $f_{\rm GW}=10^{-4}\,\mathrm{Hz}$\\
$r_{\out}$ & maximum radius of simulation & $\min(10^7R_{\rm{S}},r_{\BH,\out})$
\end{tabular}
\end{ruledtabular}
\end{table*}

\subsection{Initial Setup}
\label{sec:ini}

We consider SMBHs with masses in the range
$M_{\SMBH}=10^5$--$10^7\,M_{\odot}$, which is the mass range most
relevant for mHz-band EMRI observations. Around each SMBH, we place a
geometrically thin gaseous AGN disk and a surrounding population of
pre-existing sBHs. The initial sBH population is specified by a mass
function, a radial distribution, a velocity/inclination distribution,
and a pre-existing binary fraction. The fiducial values of the BH population parameters used throughout this work are summarized in \tabref{tab:para}.

Following \cite{xue2025determines}, the initial sBH masses are drawn from a power-law mass function with index $\beta_{\IMF}$ between $M_{\BH,\min}$ and $M_{\BH,\max}$. The initial radial distribution follows $dN/dr \propto r^{\gamma_{\rho}}$ and extends from the inner simulation boundary to $r_{\BH,\out}=GM_{\SMBH}/\sigma^2$, where $\sigma$ is the velocity dispersion of the nuclear stellar environment. The total number of pre-existing sBHs is normalized as $N_{\BH,\ini}=M_{\SMBH}/(20\Msun)$. A fraction $f_{\preexisit}$ of the initial sBH population is assigned as pre-existing binaries, with binary separations drawn up to a maximum value $R_{\max}$.

The initial velocity distribution determines how efficiently sBHs
intersect and are captured by the AGN disk. We adopt the fiducial
Gaussian inclination model of \cite{xue2025determines}, controlled by
the parameter $\beta_v$. Objects with initially inclined orbits cross
the disk only for a fraction of their orbital period. Gas drag then
damps their inclinations, allowing some sBHs to become embedded in the
disk. Once embedded, sBHs can migrate, accrete gas, form binaries,
undergo binary--single interactions, and participate in hierarchical
mergers.

The inner boundary of the simulation is set to $r_{\LISA}$, defined as
the orbital radius corresponding to a gravitational-wave frequency
$f_{\rm GW}=10^{-4}\,\mathrm{Hz}$. When an sBH reaches this radius while
remaining bound to the SMBH, we classify it as a wet EMRI candidate.
The outer boundary is set to
$r_{\out}=\min(10^7R_{\rm S},r_{\BH,\out})$, so that the simulation
covers the region where the AGN disk and the initial sBH distribution
overlap.

\subsection{AGN Disk Properties\label{sec:AGN}}

We compute the AGN disk
structure using the \texttt{pAGN} package \citep{gangardt2024pagn} for each SMBH mass and Eddington ratio.
The disk follows the Sirko--Goodman self-gravitating model
\citep{sirko2003spectral}. This model provides the radial profiles of
the gas density $\rho(r)$, surface density $\Sigma(r)$, aspect ratio
$h/r$, sound speed, and mid-plane temperature.

The Eddington ratio is defined as
\begin{equation}
l_E = \frac{L}{L_{\Edd}}
    = \frac{\dot{M}}{L_{\Edd}/(\epsilon c^2)} ,
\end{equation}
where $\dot{M}$ is the SMBH accretion rate and $\epsilon$ is the SMBH
radiative efficiency. Unless otherwise stated, we adopt
$\epsilon=0.1$. The AGN disk extends from $5R_{\rm S}$ to
$10^7R_{\rm S}$ in the default \texttt{pAGN} setup, while the actual
simulation domain is restricted to the radial range defined in
\secref{sec:ini}.

\subsection{Physical Processes}
\label{sec:process}

The physical processes included in our model are inclination damping, radial migration, GW orbital decay, gas accretion, binary formation, and mergers. Most
prescriptions follow the fiducial model of \cite{xue2025determines}. The main update is the inclination-damping prescription, for which we adopt recent hydrodynamical results \citep{whitehead2025hydrodynamic,rowan2025hydrodynamic}. This process
controls the capture of initially inclined sBHs into the AGN disk, while
radial migration and GW orbital decay determine whether embedded sBHs can reach the LISA band before the disk dissipates. Other processes, including gas accretion, binary formation, binary--single interactions, and gas-assisted hardening, are treated as in
\cite{xue2025determines}, unless otherwise stated.

Inclined sBHs interact with the AGN disk mainly during disk crossings.
These repeated crossings dissipate orbital energy and angular momentum,
thereby damping the inclination and allowing some objects to become
embedded in the disk. We update the inclination damping rate following \cite{whitehead2025hydrodynamic,rowan2025hydrodynamic}:
\begin{equation}
\log\left(\frac{\Delta i}{i}\right)=
    \begin{cases}
        a_3\log(\tilde{m})+b_3\log(\tilde{i}_c)+c_3 &\tilde{i}<\tilde{i}_c\\
        a_3\log(\tilde{m})+b_3\log(\tilde{i})+c_3 &\tilde{i}>\tilde{i}_c
    \end{cases},\label{eq:inc_dec_2}
\end{equation}
where $\tilde{m}=2\pi\rHill^2 \Sigma_{\gas}/M_{\BH}$ and $\tilde{i}=\sin i/(h/r)$. The fitted constants are $a_3=0.67$, $b_3=-2.64$, $c_3=1.80$, and $\tilde{i}_c=4.6$. The effective inclination change rate of a sBH outside the disk is 
\begin{equation}
   \frac{di}{dt}=\frac{\Delta i}{\pi r/\vkep}. \label{eq:inc_dec_1}
\end{equation}

The velocity of an inclined sBH relative to the disk is damped on the rate
\begin{eqnarray}
\Gamma_{\inc}=-\frac{1}{v}\frac{dv}{dt}=\left(\frac{di}{dt}\right)\frac{1}{2\tan(i/2)},
\end{eqnarray}
 where $\frac{di}{dt}$ is given by \equref{eq:inc_dec_2}. In updating the sBH velocity, $\Gamma_{\inc}$ replaces the term $p_{\disk}(\Gamma_{\acc} + \Gamma_{\GDF,v})$ used in \cite{xue2025determines}. To more accurately capture the inclination damping of individual sBHs, we include $\Gamma_{\inc}$ in the timestep calculation when the sBH is located outside the disk. Inside the disk, however, we neglect $\Gamma_{\inc}$ since it introduces a large, nearly constant damping rate; the inclination and velocity are then updated following the procedure described in \cite{xue2025determines}.

For a sBH embedded in the AGN disk or crossing the disk, gas torques drive radial
migration. We write the migration rate as
\begin{eqnarray}
\Gamma_{\mig}
=&& -\frac{(dr/dt)_{\mig}}{r} \nonumber \\
= 
&& 2 f_{\mig}
\left(\frac{M_{\BH}}{M_{\SMBH}}\right)
\left(\frac{\Sigma_{\gas} \vkep r}{M_{\SMBH}}\right)
\left(\frac{h}{r}\right)^{-2}, \label{eq:mig}
\end{eqnarray}
where $f_{\mig}=2$ is adopted as an approximation and $\Sigma_{\gas}$ is the surface gas density considering the gap opening when the sBH is embedded in the disk \cite{kanagawa2018radial}, given by:
\begin{equation}
\Sigma_{\gas}=\begin{cases}
\Sigma /(1+0.04K)&\text{BH resides in the disk}\\
\Sigma &\text{BH is outside the disk}
\end{cases},\label{eq:rho_gas}
\end{equation}
where the dimensionless factor $K$ considering the local disk angular momentum transport is given by
\begin{equation}
K=\left(\frac{M_{\BH}}{M_{\SMBH}}\right)^2\left(\frac{h}{r}\right)^{-5}\alpha^{-1}.\label{eq:K_factor}
\end{equation}
When the sBH is outside the AGN disk, migration should account for the fact that the disk-crossing time constitutes only a small fraction of the orbital period, parameterized by $p_{\disk}$:
\begin{equation}
p_{\disk}=
\frac{2}{\pi}
\arcsin\left[
\left(\frac{h}{r}\right)\Big/\sin i
\right],
\end{equation}
where $i$ is the orbital inclination. The effective orbital shrinking rate due to migration is therefore reduced to $\Gamma_{\mig}p_{\disk}$.

In addition to migration due to gas torques, GW emission also contributes to orbital contraction. We adopt the leading-order expression for the GW-induced orbital decay rate following \cite{peters1964gravitational} as
\begin{eqnarray}
\Gamma_{\GW,r}
&=& -\frac{(dr/dt)_{\GW}}{r} \nonumber \\
&=& \frac{64}{5}
\frac{G^3 M_{\SMBH} M_{\BH} (M_{\SMBH} + M_{\BH})}{c^5 r^4}. \label{eq:gw}
\end{eqnarray}

\subsection{Statistical Background Evolution \label{sec:back}}

We adopt the background-function method developed in
\cite{xue2025determines} to describe the evolution of the
disk-embedded sBH population. Instead of evolving all sBHs in one function,
we separate the background into single and binary components, denoted by
$\Sigma_{\single}(r,M)$ and $\Sigma_{\binary}(r,M)$, respectively. Both components obey the evolution equation
\begin{widetext}
\begin{equation}
    \frac{1}{r}\left[\frac{\partial (r\Sigma(t,r,M))}{\partial t}+\frac{\partial(r \Sigma(t,r,M))}{\partial r}\frac{dr}{dt}+\frac{\partial(r \Sigma(t,r,M))}{\partial M}\frac{dM}{dt}\right]=F_{\rm{fall}}(t,r,M)+F_{\rm{convert}}(t,r,M),\label{eq:Back_evo}
\end{equation}
\end{widetext}
where the radial and mass evolution of sBHs is given by $\frac{dr}{dt} = -(\Gamma_{\mig} + \Gamma_{\GW,r}) r$ and $\frac{dM}{dt} = \min(\Gamma_{\acc} M, \dot{M}_{\rm Edd})$, accounting for radial migration and mass growth due to accretion. The factor $F_{\rm fall}(t, r, M)$ represents the rate at which sBHs are captured into the disk, calculated using accretion rate and inclination damping rate \equref{eq:inc_dec_2} and \equref{eq:inc_dec_1}, while $F_{\rm convert}(t, r, M)$ describes the rate of conversions from single sBHs to binary sBHs due to binary formation and from binary sBHs to single sBHs via merger.

For the factor $F_{\rm convert}(t,r,M)$, we estimate the rates of conversions between single and binary sBHs as

\begin{eqnarray}
   && F_{\rm{convert},\single}=-\Gamma_{\capt}\Sigma_{\single}+\Gamma_{\GDF,s}\Sigma_{\binary}\\
    &&F_{\rm{convert},\binary}=\frac{1}{2}\Gamma_{\capt}\tilde{\Sigma}_{\single}-\Gamma_{\GDF,s}\Sigma_{\binary}
\end{eqnarray}
where $\tilde{\Sigma}_{\single}(r,M) = \Sigma_{\single}(r, M - \bar{m}_{\DBH})$ is shifted under the assumption that a single sBH forms a binary with an average mass $\bar{m}_{\DBH}(r)$. The “merger rate” from binaries to singles, $\Gamma_{\GDF,s}$, is estimated from gas dynamical friction (the dominant hardening effect) at the separation of Hill radius, where most binaries are formed. This estimation assumes no mass loss during the merger to simplify the process. The gas capture rate, $\Gamma_{\capt}$, defined as the rate of binary formation driven by gas dynamical friction in the AGN disk (equation (E40) in \cite{xue2025determines}), is estimated assuming the relative velocity between the single sBHs is dominated by the shear velocity $v_{\rm shear}$, and a uniform random factor $p_{\rm uni} = 0.5$ is applied to account for uncertainties in capture probability.

\subsection{Monte Carlo Simulation\label{sec:individual}}

In the binary formation process described in \cite{xue2025determines}, a single sBH can only form a binary with another single sBH. Accordingly, we replace all stellar properties $(n_{\DBH}, \bar{m}_{\DBH}, h_{\DBH}, \sigma_{\DBH})$ in the binary formation formula with their single-sBH counterparts, calculated from the single-sBH surface density distribution $\Sigma_{\single}(r, M)$.

When binary formation and mergers are incorporated into the background function, the weight assigned to each individual sBH sample (single or binary) at generation $g$ is updated as
\begin{equation}
w = \left( \frac{1}{2} \right)^{\max(g-1,0)}\frac{1}{N_{\AGN}}.\label{eq:weight}
\end{equation}
We assign generation numbers such that pre-existing sBH binaries have $g=0$ and pre-existing single sBHs have $g=1$, with the remnant’s generation increased by one after each merger. Because our simulations track only one sBH at a time, we correct for the fact that its merger partner may already be of higher generation. This is handled through a weighting scheme: each merger is simulated independently by all of its first-generation progenitors (or zeroth if they are pre-existing binaries), and the weights of all simulated branches sum to unity. For example, a third-generation remnant formed from two $2\mathrm{g}$ sBHs is simulated four times (each with weight $1/4$), while a remnant formed from a $1\mathrm{g}$ and a $2\mathrm{g}$ progenitor is simulated three times with weights $(1/4,\,1/4,\,1/2)$. This procedure ensures that every physical merger contributes exactly one effective event, despite not explicitly tracking the companion’s prior merger history.

The simulation boundaries are defined as
\begin{eqnarray}
r_{\out} &=& \min(r_{\disk,\out}, r_{\BH,\out}),\\
r_{\rmin} &=& \max(r_{\disk,\rmin}, r_{\LISA,\out}),
\end{eqnarray}
where $r_{\disk,\out} = 10^7 R_s$ is the outer boundary of the AGN disk in the \cite{sirko2003spectral} model, generated using the \texttt{pAGN} package \cite{gangardt2024pagn}. The parameter $r_{\BH,\out}$ denotes the maximum orbital radius of sBHs from the SMBH, adopted from \cite{xue2025determines}. When initializing the sBH samples at $t = 0$, we set the minimum sBH radius to $r_{\BH,\rmin} = r_{\LISA,\out}$, assuming that no sBHs are initially within the LISA band.

\subsection{LISA Detectability\label{sec:SNR}}

We estimate the signal-to-noise ratio (SNR) using the \texttt{FastEMRIWaveforms (FEW)} framework \cite{Katz_FastEMRIWaveforms,chua2021rapid,katz2021fast,speri2024fast,chapman2025fast}, which is part of the \texttt{Black Hole Perturbation Toolkit}. The SNR is computed as a noise-weighted inner product in the frequency domain,
\begin{equation}
\mathrm{SNR}^2(e,\vec{a},i) =4 \int \frac{|h_c(f)|^2}{S_n(f)}\,df,
\label{eq:SNR}
\end{equation}
where $S_n(f)$ represents the sky-averaged LISA sensitivity curve\cite{babak2017science,amaro2017laser}. The strain $h_c(f)$ is evaluated using the Kerr-eccentric waveform model under the assumptions of zero eccentricity ($e = 0$), zero inclination ($X = \cos i = 1$), and a dimensionless SMBH spin parameter $a = 0.9$, aligned with the rotation axis of the disk. The strain is further averaged over all sky orientations.

We neglect the influence of gas interactions on the binary trajectory, as these effects are typically subdominant in the LISA frequency range compared to the leading-order gravitational radiation.

A detection threshold of $\mathrm{SNR} \ge 20$ is adopted for LISA-detectable EMRIs. The \texttt{FEW} framework used for the SNR estimation in \equref{eq:SNR} does not include Galactic confusion noise, which is expected to reduce the detectability of low-frequency EMRIs and therefore likely causes our detection rates to be overestimated. Since the strain amplitude scales inversely with luminosity distance, we estimate the maximum detectable redshift $z_{\max}(M_{\SMBH}, M_{\BH})$ for EMRIs within AGN disks hosting SMBHs of mass $\log(M_{\SMBH}/\Msun) \in [5, 7]$, neglecting gas interactions. This redshift limit is then used to infer the global EMRI detection rate expected for LISA.

\subsection{AGN Demography\label{sec:cos}}

In the AGN disk model of \cite{sirko2003spectral}, the SMBH mass $M_{\SMBH}$ and the Eddington ratio $l_E$ are the primary factor governing variations in the gas density and aspect ratio. Following \cite{ananna2022bass}, we assume that the SMBH mass and Eddington ratio are independent. Consequently, they critically influences the migration rates of embedded sBHs, as well as their merger rates and resulting mass distributions \cite{xue2025determines}. 

We adopt the Eddington Ratio Distribution Function (ERDF) from \cite{ananna2022bass}
\begin{equation}
\label{eq:Edd}
   \xi_1(\log l_E)\equiv\frac{dN}{d\log l_E}\propto \left[\left(\frac{l_E}{l_E^*}\right)^{-0.02}+\left(\frac{l_E}{l_E^*}\right)^{2.04}\right]^{-1},
\end{equation}
where $l_E^* = 10^{-1.19} = 6.46 \times 10^{-2}$.

We also consider a lognormal ERDF motivated by \cite{kollmeier2006black},
\begin{equation}
    \xi_2(\log l_E)=\mathcal{N}(\log l_E\,|\,\log(0.25),\,0.3^2).
\end{equation}
We consider the Eddington ratios $l_E$ in the range $10^{-3} \le l_E \le 10^{0}$ and neglect super-Eddington AGNs, as they represent a small fraction of the AGN population.

\begin{figure*}[ht]
    \centering
    \includegraphics[width=1.0\linewidth]{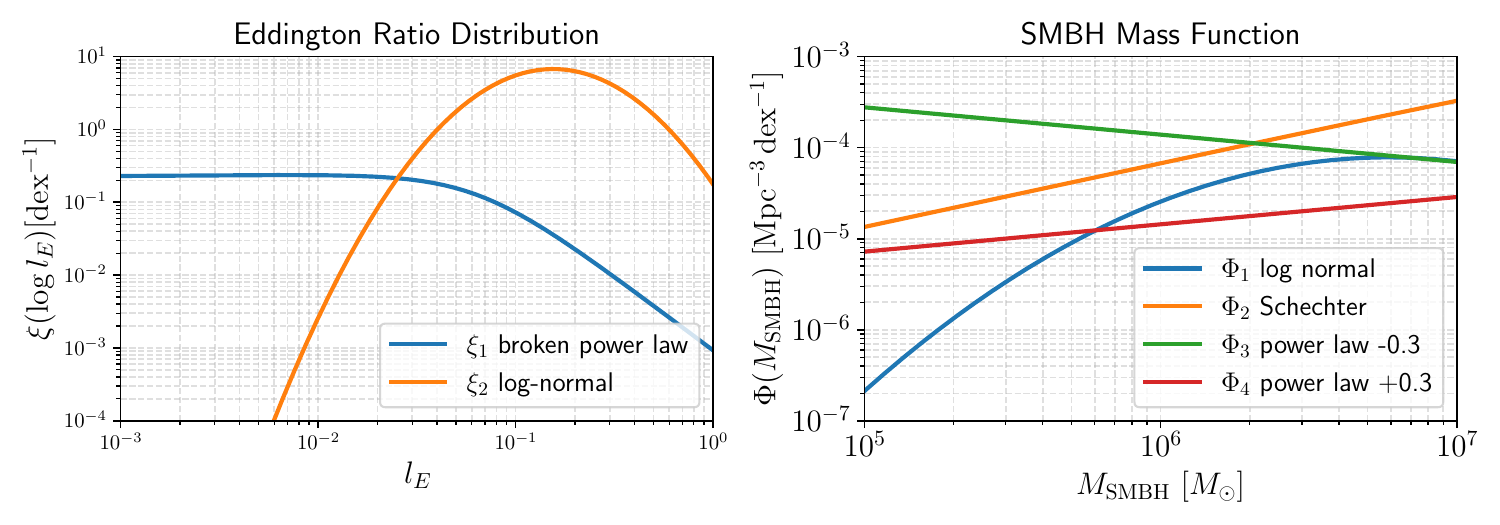}
    \caption{Left: Eddington ratio distribution functions. The broken power law $\xi_1$ (blue) favors low Eddington ratios, while the log-normal $\xi_2$ (orange) peaks at higher $l_E$. Right: SMBH mass functions. The $-0.3$ power law (green) predicts the most low-mass AGNs, the log-normal $\Phi_1$ (blue) the fewest, with the Schechter $\Phi_2$ (orange) and $+0.3$ power law (red) in between.}
    \label{fig:Mass_func}
\end{figure*}

The ERDF $\xi_1$ implies that most AGNs have relatively low Eddington ratios, $\log l_E \in [-3, -1]$, with a rapid decline at higher values. In contrast, the ERDF $\xi_2$ peaks at $l_E \approx 0.25$, as shown in the left panel of \figref{fig:Mass_func}.

The SMBH mass function of AGN $\Phi(M_{\SMBH})\equiv dn_{\AGN}/d\log M_{\SMBH}$ remains highly uncertain, particularly for low-mass SMBHs and low Eddington ratios. To explore the impact of this uncertainty, we consider several representative AGN mass distribution models.

The first model is adopted from \cite{tagawa2020formation,bartos2017rapid}, following a log normal form motivated by \cite{greene2007mass,greene2009erratum,shankar2008self}:
\begin{eqnarray}
    \Phi_1(M_{\SMBH})=&&7.83\times10^{-5}\,\rm{Mpc}^{-3}\nonumber\\
     &&\times 10^{-\left(\log\frac{M_{\SMBH}}{\Msun}-6.77\right)^2/1.22}.\label{eq:Phi_1}
\end{eqnarray}

The second model is a Schechter like function taken from \cite{graham2007millennium}, which is originally applicable for $\log M_{\rm SMBH} \in [6.5,10]$; here, we extend it to cover the full mass range considered in our simulations:
\begin{eqnarray}
    \Phi_2(M_{\SMBH})&=&\ln(10)\Phi^*\left(\frac{M_{\SMBH}}{M^*_{\SMBH}}\right)^{\alpha+1}\nonumber\\
    &&\quad\cdot\exp\left[-\left(\frac{M_{\SMBH}}{M^*_{\SMBH}}\right)\right],\label{eq:Phi_2}
\end{eqnarray}
where $\log (\Phi^*/h_{70}^3)=-2.81$, $\log( M^*{}_{\SMBH}/M_{\odot})=8.46$, $\alpha=-0.30$. 

The third and fourth models are simple power law models adopted from \cite{pan2021wet}, assuming an AGN fraction $f_{\rm AGN} = 0.01$:
\begin{eqnarray}
    \Phi_3(M_{\SMBH})&=&10^{-4}\,\rm{Mpc}^{-3}\left(\frac{M_{\SMBH}}{3\times10^6\Msun}\right)^{-0.3},\label{eq:Phi_3}\\
    \Phi_4(M_{\SMBH})&=&2\times10^{-5}\,\rm{Mpc}^{-3}\left(\frac{M_{\SMBH}}{3\times10^6\Msun}\right)^{+0.3}.\label{eq:Phi_4}
\end{eqnarray}
These four models are shown in the right panel of \figref{fig:Mass_func}, exhibiting substantial differences at the low-mass end of the SMBH spectrum and highlighting the significant uncertainties in AGN demographics in this regime.

We impose a lower cutoff on the Eddington ratio at $l_E = 10^{-3}$, below which AGNs are generally considered radiatively inefficient or advection-dominated \cite{narayan1994advection,esin1997advection,ho2008nuclear,yuan2014hot}. In this regime, the radiative efficiency drops sharply ($\epsilon \ll 0.1$) and the disk becomes geometrically thick \cite{narayan1998advection,yuan2014hot}. The transition region at $10^{-3} \lesssim l_E \lesssim 10^{-2}$ marks the onset of disk truncation, which significantly slows down migration \cite{ho2008nuclear,esin1997advection}. In this work, we include this transitional regime but assume a standard thin-disk structure throughout, since truncation may occur at $10^2 R_{\rm{S}}$ to $6R_{\rm{S}}$, which is small enough compared to the LISA detection \cite{liu2001truncation,nemmen2014spectral,yuan2014hot}. Substantial uncertainties also remain regarding the properties and demographics of low-mass, low-luminosity AGNs, which are observationally difficult to constrain \cite{kormendy2013coevolution}.

The age distribution and lifetime of AGN disks remain poorly constrained. We therefore assume that ages of AGNs hosting detectable EMRIs are uniformly distributed between $0$ and $100\,\mathrm{Myr}$, and we use the time-averaged EMRI formation rate to compute the expected detection (see \equref{eq:rate}).

\subsection{EMRI Formation Rate and Detection Rate\label{sec:formation}}

The EMRI formation rate for a single AGN, $\Gamma_{\EMRI}(M_{\SMBH}, l_E)$, is defined as
\begin{eqnarray}
\label{eq:theo_rate}
&&\Gamma_{\EMRI}(M_{\SMBH},l_E)=\nonumber\\
    &&\frac{1}{t_{\AGN}}\int d\log M_{\BH}\frac{dN_{\rm{form}}( M_{\SMBH}, M_{\BH},l_E)}{d\log M_{\BH}}    \label{eq:rate}
\end{eqnarray}
where $t_{\AGN}$ denotes the AGN lifetime, which is assumed to be 100 Myr. In this way, we effectively assume that AGN ages are uniformly distributed between 0 and 100 Myr at all redshifts.

The differential EMRI detection rate, accounting for the cosmological distribution of AGNs, is given by
\begin{eqnarray}
    &&\frac{d^4\Gamma_{\rm{detect}}}{d\log M_{\SMBH}\,d\log M_{\BH}\, dz\, d\log l_E}=\nonumber\\
    &&\quad\quad\quad\quad\frac{d^2n_{\AGN}}{d\log M_{\SMBH}d\log l_E}\cdot\frac{d V_c}{(1+z)dz}\nonumber\\
    &&\quad\quad\quad\quad\cdot\frac{dN_{\rm{form}}( M_{\SMBH}, M_{\BH},l_E)}{d\log M_{\BH}}\frac{\Theta(z_{\max}-z)}{t_{\AGN}}.\nonumber\\
    \label{eq:detect}
\end{eqnarray}
where $V_c$ is the comoving volume element, and the Heaviside step function $\Theta(z_{\max} - z)$ ensures that only EMRIs detectable at redshift $z$ are counted. Here, $z_{\max}(M_{\SMBH}, M_{\BH})$ is the maximum observable redshift for an EMRI with a given SMBH and sBH mass combination, computed via the luminosity distance corresponding to the SNR detection threshold described in \secref{sec:SNR}. 

We adopt a flat $\Lambda$CDM cosmology with parameters $\Omega_M = 0.3158$, $\Omega_\Lambda = 1 - \Omega_M$, and $h = 0.6732$, corresponding to $H_0 = 100h~\mathrm{km\;s^{-1}\,Mpc^{-1}}$ \cite{aghanim2020planck}.

\section{Results\label{sec:result}}

\begin{figure*}
    \centering
    \includegraphics[width=1.0\linewidth]{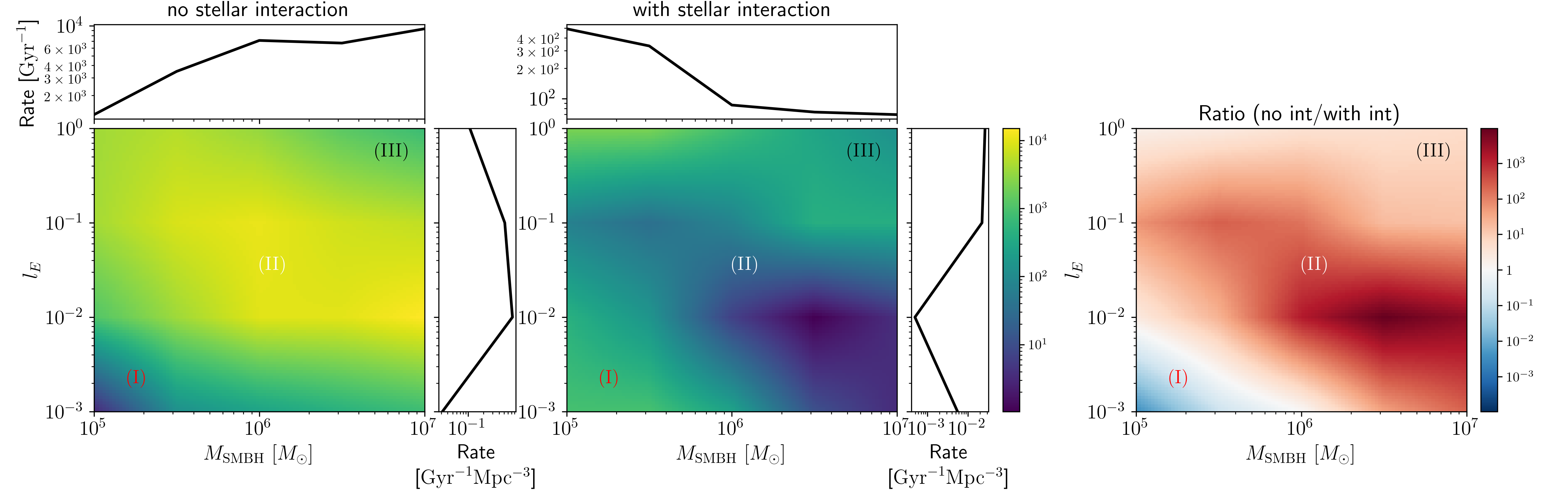}
    \caption{EMRI formation rates $\Gamma_{\EMRI}$ as a function of SMBH mass and Eddington ratio, for models without (left) and with (middle) stellar interactions. Data points are listed in \tabref{tab:theo}, and the surfaces are smoothed using a \texttt{RegularGridInterpolator} in the \texttt{SciPy} package. The top panels show the formation rates weighted by the Eddington ratio distribution $\xi_1$ given by \equref{eq:Edd}, while the side panels show the corresponding local rates integrated over the x-axis using SMBH mass function $\Phi_1$ given by \equref{eq:Phi_1}. The right panel shows the ratio of the EMRI formation rate without stellar interactions to that with stellar interactions.}
    \label{fig:form_rate}
\end{figure*}

\begin{table*}
\caption{\label{tab:theo}%
The EMRI formation rate $\Gamma_{\EMRI}(M_{\rm SMBH}, l_E)$, defined in \equref{eq:theo_rate}, evaluated across SMBH mass $M_{\rm SMBH}$ and Eddington ratio $l_E$.  Results are shown for two models: model without stellar interactions and model that includes stellar dynamical interactions.
}

\begin{ruledtabular}
\begin{tabular}{cccc}
SMBH mass& Eddington ratio&   \multicolumn{2}{c}{$\Gamma_{\EMRI}$ ($\rm{Gyr^{-1}}$)} \\
$M_{\SMBH}[\Msun]$& $\log l_E$&   no stellar interaction  & with stellar interaction \\
\hline
$10^5$&$-3$&3&878\\
  &$-2$&1116&455 \\
    &$-1$&4010&66 \\
      &$0$&3845&2286  \\
 \hline
  $10^{5.5}$&$-3$&124&916 \\
  &$-2$ &3729&152  \\
    &$-1$&  8507& 35  \\
      &$0$& 5186 & 2258 \\
 \hline
  $10^6$&$-3$& 257 & 263         \\
  &$-2$  &  9621& 6.5   \\
    &$-1$&  10995&75   \\
      &$0$& 3227 &793  \\
 \hline
  $10^{6.5}$&$-3$&447 &11 \\
  &$-2$&  9843&1 \\
    &$-1$&  7474&417 \\
      &$0$& 1309&286 \\
 \hline
  $10^7$&$-3$& 725&3  \\
  &$-2$& 15003 &4   \\
    &$-1$& 6025 &397    \\
      &$0$ & 707 &126\\
\end{tabular}
\end{ruledtabular}
\end{table*}

\subsection{EMRI Formation in AGNs}

The rate of EMRI formation for different SMBH masses and Eddington ratios is summarized in \tabref{tab:theo} and plotted in \figref{fig:form_rate}. We present the EMRI formation rates for two models: one excluding stellar dynamical interactions (left panel) and one including them (right panel). The key distinction is that, in the absence of interactions, sBHs embedded in the AGN disk do not form binaries; consequently, only pre-existing binaries can be temporarily lifted out of the disk due to the merger recoils. This non-interaction model isolates the effects of merger and  binary--single interaction recoil kicks on EMRI formation.

Relative to the no-interaction case, we classify the changes in EMRI formation into three regimes: (I) low SMBH mass and low Eddington ratio AGNs ($M_{\SMBH} \lesssim 10^{6}\,\Msun$ and $l_E \lesssim 0.01$, bottom-left of the panels in \figref{fig:form_rate}), where the EMRI formation rate is enhanced; (II) intermediate (“normal”) AGNs, primarily along the diagonal from the top-left to the bottom-right of the panels \figref{fig:form_rate}, where stellar dynamical interactions lead to a significant reduction in the EMRI formation rate; and (III) high SMBH mass and high Eddington ratio AGNs ($M_{\SMBH} \gtrsim 10^{6}\,\Msun$ and $l_E \gtrsim 0.1$, top-right of \figref{fig:form_rate}), where the reduction is comparatively modest.

Across these three regions, we identify distinct patterns in the joint distribution of sBH mass ($M_{\rm BH}$) and AGN's age ($t$) at the time of EMRI detection (\figref{fig:M_t}). In region I (low-mass low-Eddington ratio, left panel), EMRIs are narrowly clustered around $M_{\rm BH} \sim 100\,\Msun$ and occur at relatively late times ($t \sim 10$-$20$ Myr). In the region II, (middle panel), corresponding to the region of strongest suppression in \figref{fig:form_rate}, the sBH mass distribution is broad and detections occur predominantly at early times ($t \lesssim 10$ Myr). Increasing either the SMBH mass or AGN luminosity shifts the mass distribution toward higher values, while the characteristic timescale remains $\lesssim 10$ Myr. By contrast, in region III (high-mass, high-Eddington ratio, right panel), the population is more moderately concentrated, with $M_{\rm BH} \sim 40$-$200\,\Msun$ and $t \lesssim 20$ Myr.

\begin{figure}[h!]
    \centering
    \includegraphics[width=0.99\linewidth]{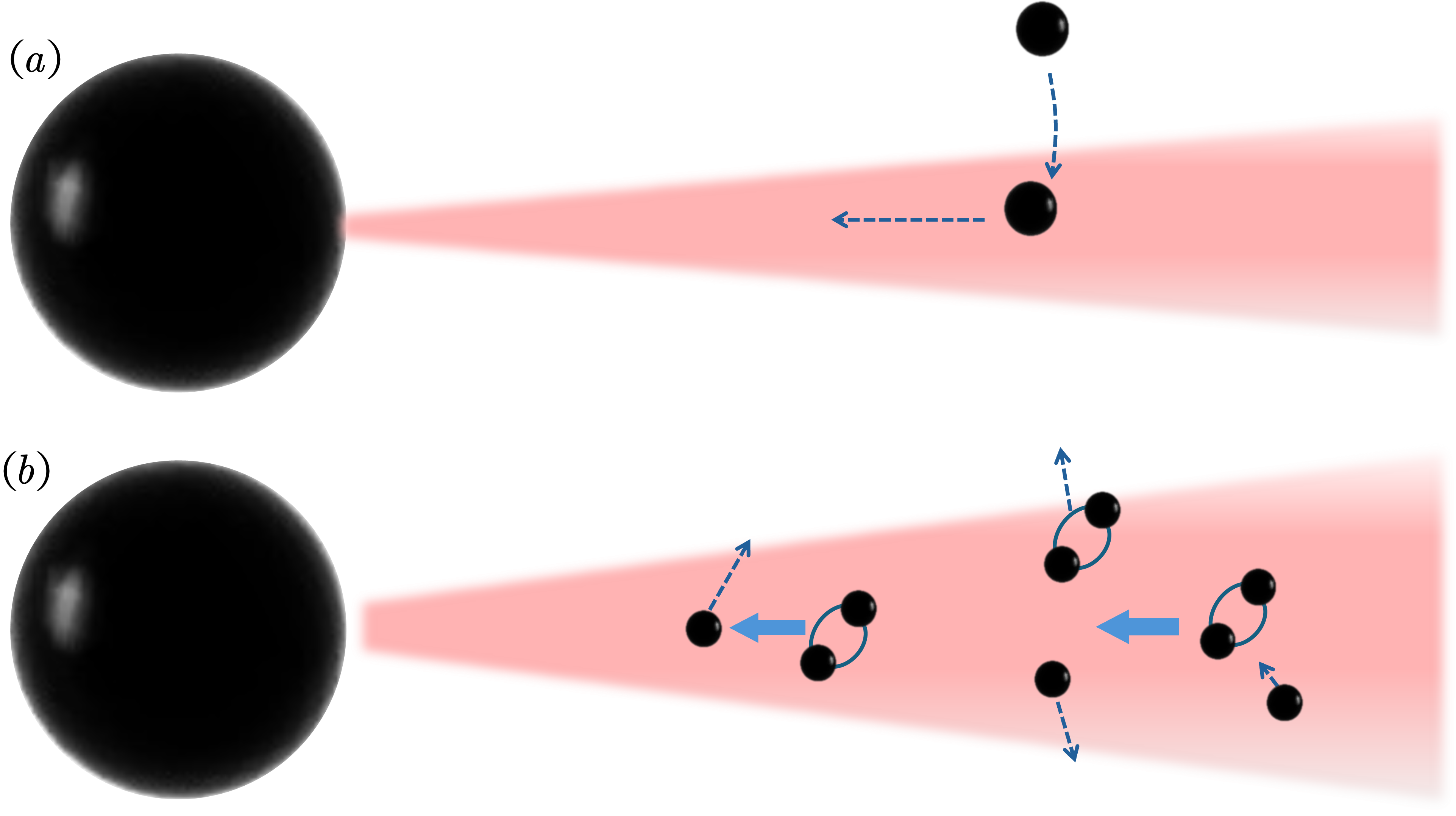}
    \caption{
{\bf Schematic illustration of mechanism affecting the EMRI formation in AGN disks. }
(a) \textit{Migration:} sBHs inclined with the AGN disk and migrate toward SMBH. 
(b) \textit{Recoil:} Recoils from merger and binary--single interaction can lift the sBHs out of the disk temporarily, slowing down inward migration, which often happens in region (II) and (III).
}
    \label{fig:agn_emri}
\end{figure}

\subsection{Physical Mechanisms Governing EMRI Formation \label{sec:physics}}

\begin{figure*}
    \centering
    \includegraphics[width=1.0\linewidth]{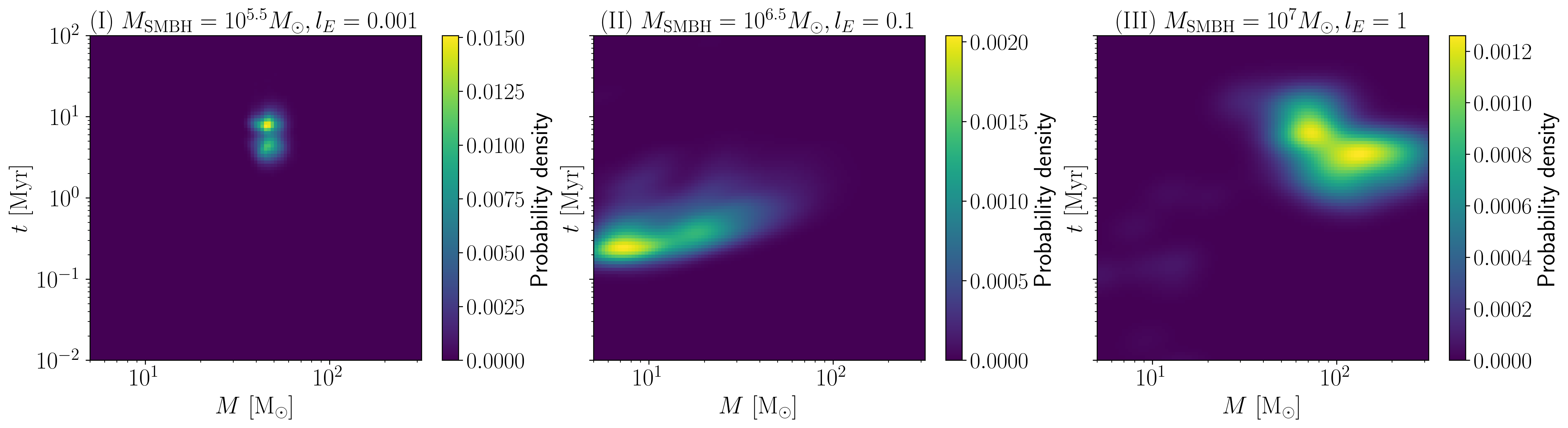}
    \caption{Smoothed and normalized distributions of secondary sBH mass ($M_{\rm BH}$) versus AGN's age at EMRI detection ($t$) for three representative AGN regimes using Gaussian KDE. Left: an example pattern of low-mass and low-luminosity AGNs (region I), with $M_{\rm SMBH}=10^{5.5}\,\Msun$ and $l_E=10^{-3}$. Middle: a representative “common’’ pattern (region II) with $M_{\rm SMBH}=10^{6.5}\,\Msun$ and $l_E=0.1$. Right: an example pattern of high-mass and high-luminosity AGN, with $M_{\rm SMBH}=10^{7}\,\Msun$ and $l_E=1$.}
    \label{fig:M_t}
\end{figure*}

\begin{figure*}
    \centering
    \includegraphics[width=1.0\linewidth]{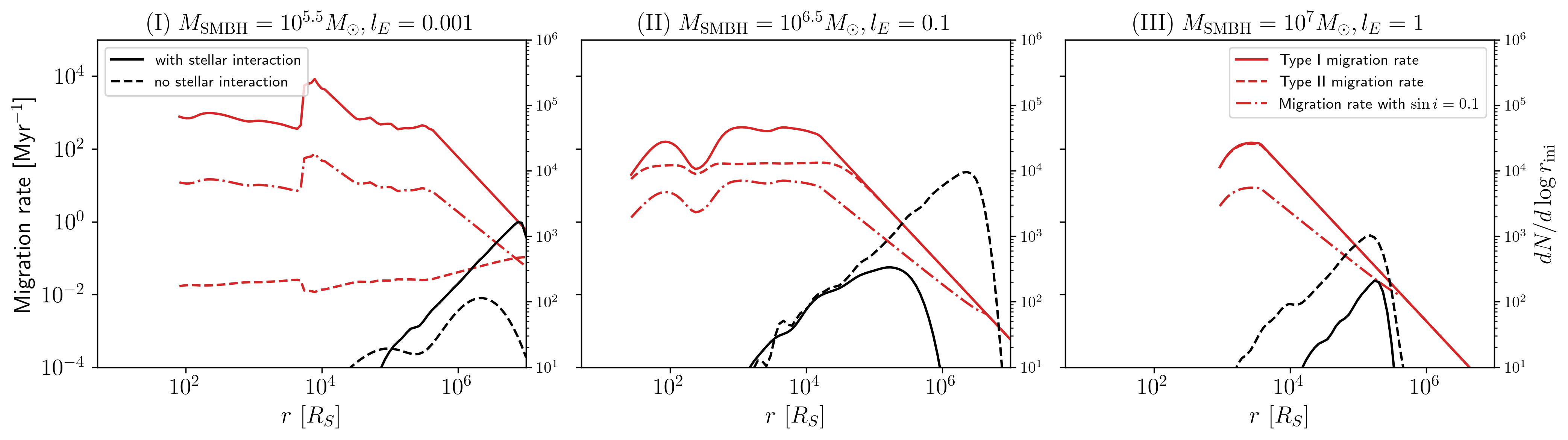}
    \caption{ Migration rates and initial sBH radii distribution of all EMRI progenitors $dN/d\log r_{\rm ini}$ for the three representative patterns shown in \figref{fig:M_t}. The solid red curve denotes the type I migration rate from equation \equref{eq:mig} with no gap opening ($K=0$), while the dashed red curve shows the type II migration rate after gap opening. The dash-dotted red curve illustrates the effective migration rate for an inclined orbit with $\sin i = 0.1$ (outside the disk). The solid and dashed black curve show the distribution of initial sBH radii of all EMRI progenitors for models with and without stellar interaction, smoothed with a Gaussian kernel. 
    }
    \label{fig:ini_rad}
\end{figure*}

We illustrate the key physical mechanisms governing EMRI formation in \figref{fig:agn_emri}. The dominant process setting the EMRI formation rate is the migration of sBHs within the AGN disk. In the absence of stellar dynamical interactions, the formation rate of EMRIs is determined primarily by two factors: (i) the supply of sBHs that become embedded in the disk, and (ii) the timescale over which they migrate into the EMRI region. The first factor depends on the total sBH population and the inclination damping rate defined in \equref{eq:inc_dec_2}, while the second is governed by type II migration given by \equref{eq:mig} for disk-embedded sBHs.

As the SMBH mass and Eddington ratio increase, the gas density rises at large radii, leading to a larger number of sBHs embedded in the disk. However, in the high-mass, high-Eddington ratio regime (region (III) of the left panel in \figref{fig:form_rate}), the migration rate drops sharply at large radii due to the increased disk aspect ratio, limiting the number of sBHs that can reach the EMRI region. This effect partially offsets the enhanced supply of embedded objects.

When stellar dynamical interactions are included, two main effects suppress EMRI formation. First, hierarchical mergers reduce the number of sBHs in the disk, while leaving the total mass of EMRIs approximately unchanged. Second, recoil kicks from mergers and  binary--single interactions can temporarily lifted sBHs out of the disk, effectively slowing their inward migration. Consistent with this picture, we find that the total mass of EMRIs is reduced by roughly an order of magnitude in most AGNs when stellar interactions are included, except in the high-mass, high-Eddington ratio regime.

Further insight is provided by \figref{fig:ini_rad}, which shows the migration rate $\Gamma_{\mig}$ (red curves) together with the initial radial distribution $dN/d\log r_{\ini}$ of EMRI progenitors at disk formation ($t=0$) for models with (solid black curves) and without (dashed black curves) stellar interactions. The type II migration rate (red dashed line) at $\Gamma_{\mig}\sim 10^{-2}\;\mathrm{Myr}^{-1}$ closely matches the largest initial radii $r_{\ini}$ in the no-interaction model (dashed black curves), where migration alone sets the baseline behavior. Deviations from this baseline in the stellar-interaction model (solid black curves) arise from hierarchical mergers and recoil-induced ejections.

In high-mass, high-Eddington ratio AGNs (region III; right panels of \figref{fig:M_t} and \figref{fig:ini_rad}), the migration speed decreases rapidly at large radii due to the large disk aspect ratio, although the gas density increases. As a result, only sBHs originating at small radii can reach the EMRI region. The total EMRI mass remains similar between the models without and with stellar interaction (not shown in the figure), indicating that hierarchical mergers dominate the modification of the formation rate. The resulting EMRIs are typically first- or second-generation remnants with relatively high masses ($\sim 100\,\Msun$), forming at intermediate times ($1\,\mathrm{Myr} \lesssim t \lesssim 20\,\mathrm{Myr}$, mostly around $10\,\mathrm{Myr}$), with recoil effects playing a secondary role.

In typical AGNs (region II; middle panels of \figref{fig:M_t} and \figref{fig:ini_rad}), the high number density of disk embedded sBH leads to frequent dynamical interactions, particularly at $t \gtrsim 10$ Myr. sBHs originating at large radii often undergo repeated cycles of temporary ejection, inclination damping, and re-capture, significantly delaying their inward migration. Because ejection at large radii leads to longer return times (due to large recoil inclination and low gas density), this effect becomes increasingly important over the AGN lifetime. 

Consequently, only a small fraction of sBHs that experience relatively few interactions can reach the EMRI region within $t \lesssim 10$ Myr. These systems typically have low generation numbers ($\lesssim 5$) and span a broad mass range ($5$-$100\,\Msun$). In this regime, hierarchical mergers do not strongly deplete the sBH population; instead, the dominant suppression arises from migration delays, leading to a reduction in the total EMRI mass by $\mathcal{O}(10)$-$\mathcal{O}(100)$ relative to the no-interaction model.

In low-mass, low-Eddington ratio AGNs (region I; left panels of \figref{fig:M_t} and \figref{fig:ini_rad}), stellar dynamical interactions can instead enhance EMRI formation. In this regime, gap opening is so efficient that the resulting type II migration rate becomes much smaller than the effective migration rate $\Gamma_{\mig}p_{\disk}$ for an sBH outside the disk with intermediate inclination, as illustrated by the dash-dotted migration track in \figref{fig:ini_rad}. In this scenario, recoil kicks accelerate their effective inward migration by temporarily ejecting disk-embedded sBHs, allowing more sBHs originating at larger radii to efficiently reach the EMRI region. Meanwhile, because only a small number of sBHs are embedded in such disks, hierarchical mergers can rapidly deplete the available seeds. The merger remnants may subsequently re-enter the disk and reach the EMRI region with relatively high masses at intermediate times ($t \sim 15$ Myr), producing a distinct EMRI formation channel in low-mass, low-Eddington ratio systems. Although hierarchical mergers significantly reduce the number of seeds, the enhanced migration efficiency of temporarily ejected sBHs ultimately leads to an overall increase in the total mass entering the EMRI region.

\subsection{EMRI Detections}

\begin{figure*}
    \centering
    \includegraphics[width=1.0\linewidth]{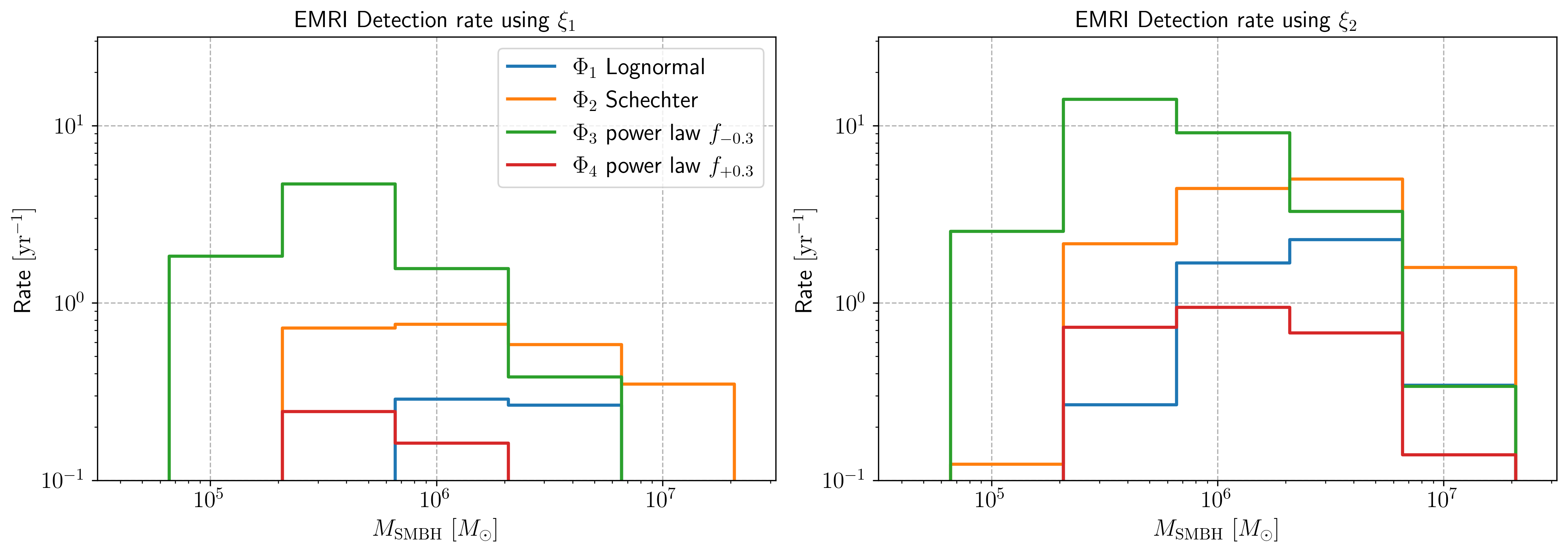}
    \caption{The expected EMRI detection rate at $z \lesssim 4$  per year per SMBH mass bin, $\Gamma_{\rm detect}$ under ERDF $\xi_1$ (left panel) and $\xi_2$ (right panel), for different SMBH mass distributions. EMRIs with primary masses of $10^{6}$-$10^{6.5}\,\Msun$ are most favorable for high-redshift detections. Both detection rates are computed using four AGN SMBH mass functions: log-normal $\Phi_1$ (red), Schechter function $\Phi_2$ (orange), and two power-law distributions, $\Phi_3$ (green) and $\Phi_4$ (blue). }
    \label{fig:N_det}
\end{figure*}
\begin{figure*}
    \centering
    \includegraphics[width=1.0\linewidth]{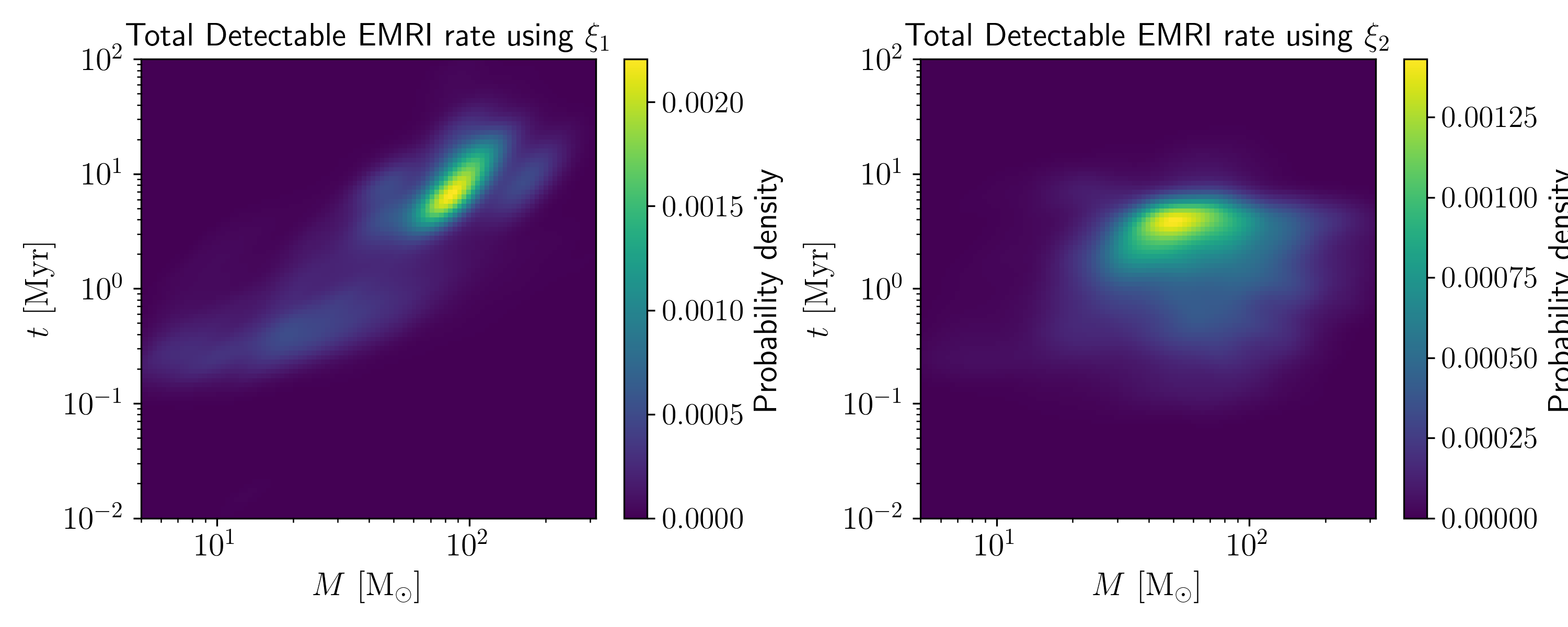}
    \caption{Same as \figref{fig:M_t} but restricted to EMRIs detected by LISA under ERDF $\xi_1$ (left panel) and $\xi_2$ (right panel). The SMBH mass distribution is calculated using the lognormal SMBH mass function $\Phi_1$ in \equref{eq:Phi_1}.}
    \label{fig:M_t_det}
\end{figure*}

\begin{table*}
\caption{\label{tab:det}%
Expected EMRI detection rates with LISA, $\Gamma_{\rm detect}$ (see \equref{eq:theo_rate}), evaluated for SMBH mass functions $\Phi_1$-$\Phi_4$ and ERDFs $\xi_1$ and $\xi_2$.}

\begin{ruledtabular}
\begin{tabular}{cc|cccc}
$\Gamma_{\rm{detect}}\;[\rm{yr}]^{-1}$& SMBH mass function& $\Phi_1(M_{\SMBH})$  & $\Phi_2(M_{\SMBH})$  & $\Phi_3(M_{\SMBH})$ & $\Phi_4(M_{\SMBH})$\\
EDRF&& log normal&Schechter-like&power law $f=-0.3$&power law $f=+0.3$\\
\hline
$\xi_1(\log l_E)$ &broken power law&0.7&2.5&8.5&0.6\\
$\xi_2(\log l_E)$ &log normal&4.6&13.3&29.3&2.6\\
\end{tabular}
\end{ruledtabular}
\end{table*}

We estimate the expected LISA detection rate of wet EMRIs at $z \lesssim 4$ under different SMBH mass functions and ERDFs, as summarized in \tabref{tab:det}. The overall detection rate ranges from $\sim 0.6$ to $29.3\ \mathrm{yr^{-1}}$. For comparison, the model that excludes stellar dynamical interactions yields significantly higher rates of $10$-$300\ \mathrm{yr^{-1}}$, particularly when adopting the lognormal ERDF $\xi_2$, broadly consistent with the estimates of \cite{pan2021wet}.

The discrepancy between our model including stellar interactions and that of \cite{pan2021wet}, which largely neglects stellar interactions, likely arises from two main factors: (i) our more detailed treatment of inclined sBHs, and (ii) the inclusion of  binary--single interactions and hierarchical sBH mergers within the AGN disk. These processes both deplete the reservoir of sBH seeds and modify migration timescales (see \tabref{tab:theo} and \secref{sec:physics}). Aside from \cite{pan2021wet}, \cite{tagawa2020formation} also estimated a local wet EMRI rate of $0.1$-$0.6\ \mathrm{Gpc^{-3}\,yr^{-1}}$, whereas our model predicts $0.025$-$0.39\ \mathrm{Gpc^{-3}\,yr^{-1}}$ for the lognormal ERDF $\xi_2$. One key difference is that our inclination damping rate given by \equref{eq:inc_dec_2} is generally smaller than the gas dynamical friction rates adopted in \cite{tagawa2020formation}, as shown in recent hydrodynamic studies \citep{whitehead2025hydrodynamic,rowan2023black}.

The distribution of SMBH masses (primary masses) for detectable EMRIs is shown in \figref{fig:N_det}. The left panel adopts the broken power-law ERDF $\xi_1$, while the right panel uses the lognormal ERDF $\xi_2$. The ERDF $\xi_1$ favors low Eddington ratios, whereas $\xi_2$ peaks near $l_E \sim 0.25$, corresponding to a higher fraction of luminous AGNs with intrinsically higher EMRI formation rates. Consequently, the overall detection rate is generally higher for $\xi_2$. The SMBH mass function also plays an important role: a higher abundance of low-mass AGNs leads to an increased number of detectable EMRIs.

A major source of systematic uncertainty arises from the poorly constrained SMBH mass function at the low-mass end, as well as the ERDF at low Eddington ratios. Low-luminosity, low-mass AGNs remain observationally difficult to identify \cite{satyapal2016challenges,sartori2015search}, yet they may contribute significantly to the local wet EMRI population. Furthermore, current observations suggest an ERDF resembling $\xi_1$ in the local universe and one similar to $\xi_2$ at higher redshift (e.g., Fig.~7 of \cite{kelly2013demographics}). Whether this trend reflects a true redshift evolution or is primarily driven by selection effects remains uncertain. In this context, future EMRI detections with LISA could provide unique constraints on the abundance and properties of faint AGNs. A similar conclusion has been drawn in the context of tidal disruption events (TDEs), where \cite{stone2016rates} showed that TDE rates are dominated by low-mass SMBHs, making them a sensitive probe of the low-mass end of the SMBH mass function.

The joint distribution of EMRI secondary mass and host AGN age is shown in \figref{fig:M_t_det}, using signal-to-noise ratios computed in \secref{sec:SNR}. Both panels assume the lognormal SMBH mass function $\Phi_1$ and are normalized to unity, with the left and right panels corresponding to ERDFs $\xi_1$ and $\xi_2$, respectively. We find that this distribution is largely insensitive to the choice of SMBH mass function. For $\xi_1$, the typical sBH masses are slightly higher than in the $\xi_2$ case, although both scenarios favor detections at relatively early AGN ages ($t \lesssim 20\ \mathrm{Myr}$). As discussed in \secref{sec:physics}, sBHs that would otherwise reach the EMRI region at later times are often delayed by temporary ejections following mergers and  binary--single interactions. When adopting $\xi_2$, the distribution becomes more tightly constrained to $t \lesssim 10\ \mathrm{Myr}$.

These results suggest that EMRIs can serve as probes of AGN evolutionary stage. In extremely high Eddington ratio AGNs hosting massive SMBHs, as well as in faint AGNs with low-mass SMBHs, EMRI detections tend to correspond to relatively older systems ($t_{\rm AGN} \sim 10$-$20$ Myr) with more massive secondary sBHs. In contrast, detections in intermediate regimes typically trace younger AGNs ($t_{\rm AGN} \lesssim 10$ Myr) with lower-mass secondaries. Very old AGNs ($t \gtrsim 30$ Myr) contribute negligibly to the wet EMRI population. Overall, EMRIs are most likely to be detected in relatively young AGNs, and joint measurements of SMBH mass, sBH mass, and merger generation may provide an independent probe of AGN lifetimes, complementing merger population studies \cite{xue2025determines,mckernan2025mcfacts}.

\subsection{Interesting Events: Binary EMRIs}
Recent studies have explored the detectability of binary EMRIs  \cite{meng2026observability,jiang2024distinguishability,sharma2026benchmark}, where a binary sBH orbiting around the SMBH. In our simulations, we also identify several intriguing but rare subclasses of such events. Although such systems may theoretically occur widely in certain AGNs, their detectability is limited by the intrinsically low EMRI detection rate.

A small fraction of EMRIs involve binary sBHs serving as the secondary object. In typical high-mass and high-luminosity AGN environments, these systems constitute only $\sim 1\%$ of all EMRIs. However, in low-mass and low-luminosity AGNs, the fraction rises to nearly $10\%$. Because the two components of the binary enter the EMRI region at roughly the same time, these systems may enable the observation of hierarchical triple interactions, or even direct binary mergers occurring within the LISA detection region, offering a powerful probe of the AGN environment \cite{han2019testing}. Although the binary separations remain safely below the Hill radius at $r_{\LISA}$, they may nonetheless experience tidal disruption if the EMRI orbital decay proceeds faster than binary hardening. In addition, Kozai-Lidov oscillations may excite large eccentricities, creating unique signals in GW detection \cite{su2025gas,naoz2016eccentric}.

\subsection{Interesting Events: Quasi-Periodic Eruptions}
In addition, interactions between sBHs and the AGN disk can generate flares that may manifest as quasi-periodic eruptions (QPEs) \cite{linial2023emri,franchini2023quasi}, although the current observational period of QPEs is normally longer than the LISA sensitivity ($10^4$ second) \cite{pasham2024alive,giustini2020x,miniutti2019nine}. In addition, \cite{zhou2024probing,zhou2024probing2} analyze existing QPE events and find that four have low eccentricities ($e < 0.1$), while one exhibits a mild eccentricity ($e \sim 0.25$).

For such electromagnetic counterparts to arise in the sBH-disk interaction scenario, the sBH must reside outside the disk plane and cross the disk twice per orbital period. This configuration may occur if the sBH is initially misaligned upon entering the AGN disk. In our model, it can also result from recoil kicks induced by  binary--single interactions or mergers, which can temporarily displace the sBH from the disk. Such recoil events can naturally produce high-eccentricity QPEs.

However, this channel faces a significant challenge: the predicted luminosities from sBH-disk collisions may be insufficient to explain the brightest observed QPEs \cite{linial2023emri,miniutti2013high,shu2017central,2025Wevers,tagawa2023flares,suzuguchi2025}. Although our current framework does not explicitly model stellar migration, the resulting distributions of sBH spatial positions and orbital inclinations nonetheless provide useful insight into the viability of this mechanism.

Within our current framework, we estimate the QPE formation rate by identifying EMRIs whose orbits remain outside the AGN disk at the characteristic LISA radius, $r_{\mathrm{LISA}}$. Interestingly, we find that a substantial fraction (exceeding 40\%) of EMRIs satisfy this criterion in the following AGN environments: $(\log M_{\SMBH}, \log l_E) = (5.5,0), (6.5,-2), (6.5,-1), (7.0,-3), (7.0,-2), (7.0,-1)$. Systems with $\log M_{\SMBH}=6.0$ yield too few EMRIs to draw firm conclusions. These regimes are broadly consistent with the properties of observed QPE host AGNs, and offer a natural physical explanation: the AGN disk must be dense enough to capture sBHs and facilitate EMRI formation, while still thin enough that recoil kicks from mergers and  binary--single interactions can eject sBHs out of the disk plane temporarily. This balance favors Eddington ratios of order $\mathcal{O}(0.1-1)$.

Despite the high fraction in these specific regimes, the local formation rate remains low, at the level of $\sim 10^{-11}\,\mathrm{Mpc^{-3}\,yr^{-1}}$, compared to the QPE local formation rate of $\sim 10^{-8}\,\mathrm{Mpc^{-3}\,yr^{-1}}$ \cite{arcodia2024cosmic}. This difference arises from the overall rarity of EMRIs and their limited occurrence in low-luminosity AGNs. However, current QPEs are likely coming from an orbit with larger radii, where stellar interactions are very frequent and the BHs are not likely to migrate into the LISA detection range, meaning our model will under-estimate the rate. Nevertheless, stars which is  far more numerous than sBHs may also offer a more promising pathway for producing  disk-crossing origin QPE events \cite{linial2025qpes}.

\subsection{Limitations and Uncertainties}

Our estimated detection rates rely on several key assumptions, including the SMBH mass function, AGN lifetime distribution, and Eddington ratio distribution. These quantities remain poorly constrained by current observations, particularly the AGN lifetime. In addition, our simulations impose an upper limit of $15\,\Msun$ on the mass of pre-existing single sBHs. In reality, alternative formation channels may produce more massive sBHs,particularly remnants of very massive stars formed in situ within the AGN disk. Such objects could increase the maximum masses reached through hierarchical mergers \cite{xue2025determines} and thereby modify the secondary mass distribution of EMRIs.

Our treatment of migration also introduces uncertainties. In particular, we model the transition between type I and type II migration when an sBH is temporarily ejected from the AGN disk. In low-mass, low-Eddington ratio AGNs, type I migration (applicable to sBHs outside the disk) can be significantly faster than type II migration within the disk (see the left panel of \figref{fig:ini_rad}). For off-disk sBHs, we adopt an effective radial migration rate of $\Gamma_{\rm mig} p_{\rm disk}$ based on the type I prescription. However, this process remains poorly understood and may introduce additional uncertainties into the EMRI formation rate.

Besides, we adopt the standard Sirko disk model \cite{sirko2003spectral}, which assumes a constant accretion rate. In contrast, alternative models (e.g., Thompson-like disks; \cite{thompson2005radiation,epstein2025time}) allow for in situ star formation at large radii, which can deplete the gas supply and reduce the accretion rate at smaller radii. These in situ-formed sBHs may provide an additional population of disk-embedded objects, enhancing our EMRI rates, although they are expected to contribute only a small fraction to hierarchical mergers \cite{xue2025determines,tagawa2020formation}. Since the in-situ sBHs typically originate at large radii and migrate inward over relatively long timescales, such objects are still likely to experience recoil kicks and dynamical interactions similar to pre-existing sBHs. We therefore expect this effect to have a limited impact on our results.

In addition, our model assumes a static AGN disk. In reality, AGNs may exhibit time-dependent (“changing-look”) behavior, and feedback from stars and embedded sBHs can also modify the disk structure \cite{epstein2025time}. Such global evolution could affect both migration and interaction rates, introducing further uncertainties. Moreover, if AGN activity is intermittent, our assumed $\sim 100$ Myr lifetime should be interpreted as an effective timescale. If the interval between active episodes is short compared to the stellar relaxation timescale, the sBH and stellar distributions will not change significantly, and our treatment remains valid. Conversely, if the quiescent intervals are long, each active episode should be treated as a new AGN phase, effectively resetting the system to $t=0$.

Another source of uncertainty arises from the possibility of migration traps, where inward and outward torques balance and halt radial migration. Such traps have been identified in both analytical models \cite{bellovary2016migration,jimenez2017improved} and hydrodynamical simulations \cite{masset2017coorbital,lyra2010orbital}, typically in high-mass, high-luminosity AGNs, although \cite{grishin2024effect} suggests they may also occur in lower-mass systems. The presence of migration traps would enhance in-disk mergers while suppressing EMRI formation by preventing sBHs from reaching the central region. If migration traps are common, the detectable EMRI rate could be reduced by a factor of $\mathcal{O}(1)$-$\mathcal{O}(10)$.

Finally, as noted by \cite{xue2025determines}, our semi-analytical framework may overestimate the rate of sBH interactions and binary formation compared to full $N$-body simulations. This suggests that the true EMRI formation rate may be somewhat higher than our estimates.

\section{Conclusion and Discussion}
\label{sec:conclusion}

In this work, we performed a suite of AGN disk simulations covering a broad range of SMBH masses ($10^5$-$10^7\,\Msun$) and Eddington ratios ($10^{-3}$-$1$). We weighted their contributions to the EMRI rates according to the observed AGN mass and luminosity distributions, including their redshift dependence, to describe their populations. 

Our simulations yield four main results. (i) The overall EMRI detection rate is estimated to be $0.6$-$29.3\;\rm{yr^{-1}}$, with strong sensitivity to the uncertain low-SMBH-mass and low-Eddington-ratio AGN populations. (ii) The contribution from AGNs hosting massive SMBHs ($M_{\SMBH} \gtrsim 10^6\;\Msun$) remains approximately constant per unit log SMBH mass, while the contribution from low-mass AGNs strongly depends on the uncertain low-mass end of the AGN mass function. (iii) EMRIs generally form within the first 20 Myr of AGN lifetime, as recoil kicks increasingly remove sBHs from the disk at later times, indicating that detectable events trace relatively young AGN populations. (iv) Finally, the secondary black hole masses of detectable EMRIs tend to be higher in AGNs with either lower SMBH masses and lower Eddington ratios or higher SMBH masses and higher Eddington ratios, reflecting more efficient hierarchical mergers in these environments.

Beyond these general trends, our simulations reveal several rare but astrophysically intriguing EMRI subclasses, though their detectability is limited by the intrinsically low EMRI rate. Only a small fraction of EMRIs involve binary secondary black holes. While these systems constitute only $\sim 1\%$ of EMRIs in typical or high-mass/high-luminosity AGNs, the fraction rises to nearly $10\%$ in low-mass, low-luminosity AGNs whose numbers could dominate in local observations but decrease significantly at high redshift due to detection sensitivity. Additionally, interactions between secondary black holes and AGN disks can generate flares resembling quasi-periodic eruptions (QPEs). Using our simulated EMRI population, we estimate a QPE formation rate of $\sim10^{-3}\;\rm{yr}^{-1}$ with preference for Eddington ratios around 0.1. Nevertheless, stars which may be far more numerous than secondary black holes may provide a more promising pathway for disk-crossing QPEs, warranting future investigation.

Finally, we note an analogy to the regulation of tidal disruption events (TDEs) in AGNs, where stellar collisions reduce the effective TDE rate \cite{kaur2025elevated}. AGNs are also potential TDE factories, with migration driving stars inward toward the SMBH. Such TDEs may be detectable prior to or alongside EMRI activity, highlighting the broader role of AGN disks as multi-messenger environments \cite{stone2016rates}.

\begin{acknowledgments}
Z.H. is grateful for support from NASA under Grants No. 80NSSC22K0822 and No. 80NSSC24K0440. 
H.T. is supported by The National Key R$\&$D Program of China (grant No.2024YFC2207700). I.B. is grateful for support by the National Science Foundation under grant No. PHY-2309024 (IB)
\end{acknowledgments}
\nocite{*}
\section*{DATA AVAILABILITY}
                          
The data that support the findings of this article are not publicly available. The data are available from the authors upon reasonable request.

\bibliographystyle{apsrev4-2}
\bibliography{LISA_AGN}

\end{document}